\documentclass[lettersize,journal]{IEEEtran}
\usepackage{amsmath,amsfonts}
\usepackage{array}
\usepackage{subfig}
\usepackage{stfloats}
\usepackage{url}
\usepackage{verbatim}
\usepackage{graphicx}
\usepackage{cite}
\usepackage{microtype}
\usepackage{doi}
\usepackage{float}
\usepackage{mathrsfs}
\usepackage{threeparttable}
\usepackage{booktabs}
\usepackage{multicol}
\usepackage{multirow}
\usepackage{tabto}
\usepackage[ruled]{algorithm2e}
\usepackage{algpseudocode}
\usepackage{makecell}
\usepackage{array}
\usepackage{diagbox}
\usepackage{adjustbox}
\usepackage{hyperref}
\usepackage{textcomp}

\begin{document}
 \title{Confidence-Driven Deep Learning Framework for Early Detection of Knee Osteoarthritis}
\author{Zhe Wang, Aladine Chetouani, Yung Hsin Chen,  Yuhua Ru, Fang Chen, Mohamed Jarraya, Fabian Bauer, Liping Zhang, Didier Hans$^*$, Rachid Jennane$^*$
\thanks{Zhe Wang, Yung Hsin Chen and Mohamed Jarraya are with Department of Radiology, Massachusetts General Hospital, Harvard Medical School, Boston, 02114, USA (e-mail: zwang78@mgh.harvard.edu; ychen4@mgh.harvard.edu; mjarraya@mgh.harvard.edu).}
\thanks{Aladine Chetouani is with L2TI Laboratory, University Sorbonne Paris Nord, Villetaneuse, 93430, France (e-mail: aladine.chetouani@univ-paris13.fr).}
\thanks{Yuhua Ru is with Jiangsu Institute of Hematology, The First Affiliated Hospital of Soochow University, Suzhou, 215006, China. (e-mail: ruyuhua@163.com).}
\thanks{Fang Chen is with department of Medical School, Henan University of Chinese Medicine, Zhengzhou, 450046, China. (e-mail: chenfangyxy@hactcm.edu.cn).}
\thanks{Fabian Bauer is with Division of Radiology, German Cancer Research Center, Heidelberg, 69120, Germany (e-mail: fabian.bauer@dkfz-heidelberg.de).}
\thanks{Liping Zhang is with Athinoula A. Martinos Centre for Biomedical Imaging, Massachusetts General Hospital, Harvard Medical School, Boston, 02114, USA. (e-mail: lzhang90@mgh.harvard.edu).}
\thanks{Didier Hans is with Nuclear Medicine Division, Geneva University Hospital, Geneva, 1205, Switzerland. (e-mail: didier.hans@chuv.ch).}
\thanks{Rachid Jennane is with IDP Institute, UMR CNRS 7013, University of Orleans, Orleans, 45067, France (e-mail: rachid.jennane@univ-orleans.fr).}}

\markboth{Journal of \LaTeX\ Class Files,~Vol.~14, No.~8, August~2021}%
{Shell \MakeLowercase{\textit{et al.}}: A Sample Article Using IEEEtran.cls for IEEE Journals}

\maketitle
\begin{abstract}
Knee Osteoarthritis (KOA) is a prevalent musculoskeletal disorder that severely impacts mobility and quality of life, particularly among older adults. Its diagnosis often relies on subjective assessments using the Kellgren-Lawrence (KL) grading system, leading to variability in clinical evaluations. To address these challenges, we propose a confidence-driven deep learning framework for early KOA detection, focusing on distinguishing KL-0 and KL-2 stages. The Siamese-based framework integrates a novel multi-level feature extraction architecture with a hybrid loss strategy. Specifically, multi-level Global Average Pooling (GAP) layers are employed to extract features from varying network depths, ensuring comprehensive feature representation, while the hybrid loss strategy partitions training samples into high-, medium-, and low-confidence subsets. Tailored loss functions are applied to improve model robustness and effectively handle uncertainty in annotations. Experimental results on the Osteoarthritis Initiative (OAI) dataset demonstrate that the proposed framework achieves competitive accuracy, sensitivity, and specificity, comparable to those of expert radiologists. Cohen’s kappa values ($\kappa$ $>$ 0.85)) confirm substantial agreement, while McNemar’s test (p $>$ 0.05) indicates no statistically significant differences between the model and radiologists. Additionally, Confidence distribution analysis reveals that the model emulates radiologists' decision-making patterns. These findings highlight the potential of the proposed approach to serve as an auxiliary diagnostic tool, enhancing early KOA detection and reducing clinical workload.
\end{abstract}


\section{Introduction}
\IEEEPARstart{K}{nee} OsteoArthritis (KOA), also referred to as degenerative osteoarthropathy, is a progressive joint disease characterized by the degeneration and damage of cartilage, narrowing of joint space, and subchondral bone reactive hyperplasia \cite{kneeoa}. It is influenced by a myriad of factors, including aging, mechanical stress, trauma, and genetic predisposition \cite{multi-factor}. KOA significantly impacts patients' quality of life, causing severe pain, limited mobility, and functional impairment, which can lead to an elevated risk of chronic comorbidities such as cardiovascular disease \cite{cardiovascular}. With the aging population on the rise, the United Nations (UN) predicts that individuals over the age of 60 will constitute more than 20\% of the global population by 2050 \cite{2050}. This demographic shift highlights the urgency of addressing KOA, especially as no definitive cure or universally recognized cause has been identified to date \cite{nocure,notclear}. Given these circumstances, early diagnosis of KOA is paramount. Timely identification enables the implementation of behavioral interventions, such as weight management and physical therapy, which can significantly delay the onset and progression of symptoms \cite{weightloss}. However, achieving early diagnosis is challenging due to the subtle and gradual manifestation of radiographic changes in the early stages of the disease.

The Kellgren-Lawrence (KL) grading system \cite{KL} is widely regarded as the standard for evaluating KOA severity using plain radiographs. Table \ref{KL_grades} outlines the five grades of KOA severity, ranging from no evident signs (KL-0) to advanced KOA (KL-4) characterized by significant joint damage. Despite its widespread use, the KL grading system is inherently subjective, relying heavily on the interpretation and judgment of healthcare providers. This subjectivity often leads to variability in assessments, as different medical professionals may provide inconsistent evaluations for the same knee X-ray \cite{shamir}. Such inconsistencies can delay early diagnosis and intervention, underscoring the need for objective, reliable, and automated diagnostic tools to support clinicians.

\begin{table}[htbp]
\centering
\caption{Description of the KL grading system}
\setlength{\tabcolsep}{1.2mm}
\begin{tabular}{lll}
\toprule
Grade &  Severity & Description\\
\midrule
KL-0 & none & no signs of osteoarthritis\\
KL-1 & doubtful & potential osteophytic lipping\\
KL-2 & minimal & certain osteophytes and potential JSN\\
KL-3 & moderate & moderate multiple osteophytes, certain JSN, and some\\
&&  sclerosis\\
KL-4 & severe & large osteophytes, certain JSN, and severe sclerosis\\
\bottomrule
\end{tabular}
\label{KL_grades}
\end{table}

\begin{figure*}
\centering  
\includegraphics[width=1\textwidth]{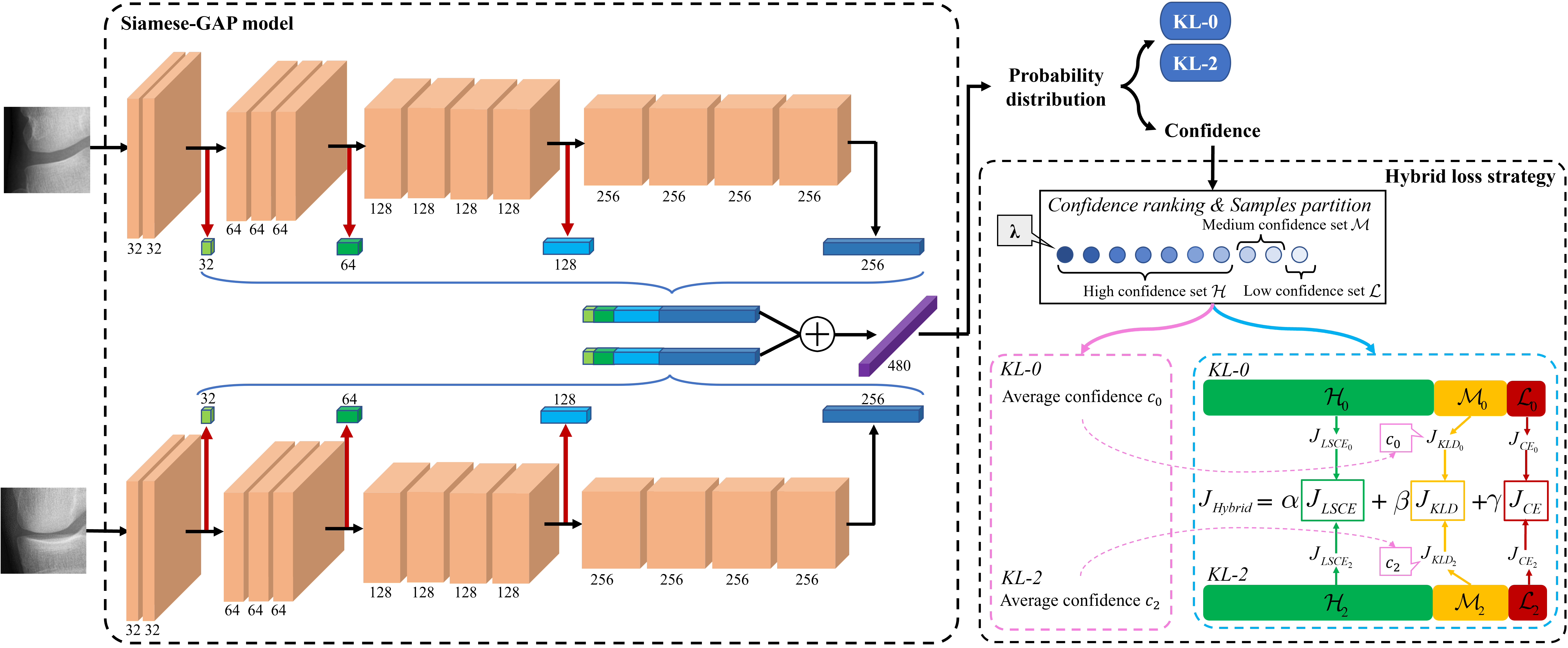}
\caption{The flowchart of the proposed approach consists of two main components: the Siamese-GAP model and the hybrid loss strategy. The overall data flow is represented by black arrows. On the left, the Siamese-GAP network includes red vertical arrows to denote GAP units, blue braces indicating concatenation operations, the $\oplus$ symbol representing element-wise addition, and a purple block for the fully connected layer followed by Softmax. On the right, the hybrid loss strategy showcases the data flow of validation and training batches with light purple and light blue arrows, respectively. Samples within each KL grade are ranked by their confidence level using $\lambda$, while the three loss functions are weighted by hyper-parameters $\alpha$, $\beta$, and $\gamma$. Additional parameters, such as $c_0$ and $c_2$, are detailed in Section \ref{hyperparaters}.}
\label{flowchart}
\end{figure*}

Recent advancements in computer hardware have significantly boosted the adoption of deep learning techniques in computer vision \cite{wang2022siamese}\cite{ 10230443}\cite{wang2024transformer}. In the context of KOA classification, several learning models have been proposed. For instance, \cite{antony2016} utilized a linear Support Vector Machine (SVM) \cite{SVM} and the Sobel horizontal image gradient to detect knee joint centers. Subsequently, they extracted Regions of Interest (ROIs) based on the recorded coordinates and fed them into a fine-tuned BVLC CaffeNet network \cite{caffe} for the classification. Similarly, \cite{antony1} employed a classical Convolutional Neural Networks (CNNs) \cite{cnn} to perform automatic KL-grade classification of knee joints. In another study, \cite{abdullah2022automatic} proposed a system combining Faster R-CNN \cite{ren2016faster} and AlexNet \cite{krizhevsky2012imagenet} with transfer learning to automatically detect the minimal joint space width (JSW) and classify the severity of KOA. The Faster R-CNN model was used for precise localization of the JSW area, while AlexNet classified KOA severity. It is noteworthy that existing methods for detecting KOA severity typically approach the problem as an image classification task, assigning each KL grade to a distinct category. However, the representation of KL grades as discrete integers is primarily a convenient abstraction, rather than a reflection of the true nature of KOA progression. Since KOA severity progresses along a continuous spectrum, KL grades are better understood as representing an ordinal scale rather than strictly categorical labels. In \cite{antony2016}, the authors treated KOA severity prediction as a regression problem, employing the Mean Square Error (MSE) loss function. This approach outperformed models using the Cross-Entropy (CE) loss function for the same task. However, single loss functions like MSE or CE do not fully capture the semi-quantitative and ordinal nature of KL grades in medical contexts. Consequently, researchers have explored alternative loss strategies to address these challenges. For example, \cite{antony1} combined CE and MSE losses with optimized weights to balance classification and regression objectives. \cite{chen} introduced a modified CE loss function that uses an adjustable ordinal matrix, assigning higher penalties for misclassifications with greater distances between predicted and true KL grades. These approaches integrate regression concepts into classification tasks, bridging the gap between the ordinal and categorical nature of KL grades and improving the interpretability and performance of KOA severity detection models. However, in addition to the continuum between different KL grades, variability also exists within the same KL grade. Medical practitioners may assign varying degrees of confidence to different samples labeled with the same KL grade. For example, two samples sharing the same KL grade label might reflect different levels of certainty in their annotations. However, due to the inherent robustness of deep learning models, samples labeled with lower certainty are either trained identically to regular-certainty samples or dismissed as noise, which can lead to models becoming overconfident \cite{overconfident}, often favoring samples with higher-confidence labels at the expense of correctly classifying lower-confidence ones. To address this issue, it is critical to develop a training strategy that adjusts the learning process based on the confidence level of each sample to allow the model to better handle uncertainty in the data. 

KOA is unequivocally diagnosed at KL-2, though it is characterized by minimal severity at this stage. Patients who progress to advanced stages (KL-3 and KL-4) often face debilitating symptoms and require invasive interventions such as total knee replacement \cite{replacement}. Consequently, the early detection of KOA is of critical clinical importance, serving as the primary focus of this study. However, KL-1 labels in the existing literature are frequently associated with substantial uncertainty, stemming from their ambiguous radiographic features that can vary significantly across interpretations. Such uncertainty introduces instability during model training, which may compromise the reliability and accuracy of predictions. Hence, our study focuses exclusively on KL-0 (no signs of KOA) and KL-2 (early definitive KOA), ensuring a clear delineation between normal and pathological conditions and avoiding the ambiguities inherent in KL-1 labels. For this, we propose a novel approach that integrates a Siamese-based learning model with an innovative hybrid loss strategy for KOA classification (KL-0 vs. KL-2). Our learning model, termed the Siamese-GAP network. Unlike traditional Siamese models, our approach also incorporates features from lower levels. To this end, we introduce a series of Global Average Pooling (GAP) layers at multiple levels of the network, enabling the extraction of critical information from each layer. These features are subsequently concatenated, ensuring a comprehensive representation that enhances KOA detection performance. Furthermore, we develop a hybrid loss strategy designed to partition the training samples in each batch into high-, medium-, and low-confidence subsets. Three distinct loss functions are applied to these subsets: Label Smoothing Cross-Entropy (LSCE), Kullback-Leibler Divergence (KLD) \cite{KLD}, and the standard Cross-Entropy (CE) loss. The final hybrid loss function is computed by optimally combining these individual loss functions with carefully tuned weights. Our proposed methodology significantly improves the accuracy and robustness of early KOA detection, providing a promising solution for clinical applications.

The primary contributions of this study are summarized as follows:
\begin{itemize} 
\item[$\bullet$] A novel Siamese-based learning framework is proposed, capable of integrating multi-level features (i.e., from low to high), facilitating comprehensive feature extraction and robust representation for KOA classification. 
\item[$\bullet$] A novel training strategy is introduced, incorporating a smart partitioning mechanism that divides each training batch into high-, medium-, and low-confidence subsets, coupled with a hybrid loss function tailored to optimize learning across varying confidence levels. 
\item[$\bullet$] Grad-CAM attention maps are employed to visualize and emphasize the key regions influencing the network's predictions, significantly enhancing model interpretability and decision transparency. 
\item[$\bullet$] The proposed global approach is evaluated from both technical and clinical standpoints, demonstrating its applicability and effectiveness in clinical practice.
\end{itemize}

\section{Proposed method}
The flowchart of our proposed global approach is presented in Fig. \ref{flowchart}. Table \ref{symbols} presents the key symbols employed in this paper.

\begin{table}[htbp]
\centering
\caption{Key symbols employed in this paper}
\setlength{\tabcolsep}{0.8mm}
\begin{tabular}{clcl}
\toprule
Symbol & Description & Symbol & Description\\
\midrule
$\mathcal{K}$ & Set of KL grades & $\mathcal{H}$ & High-confidence set\\
$\mathcal{D}$ & Overall dataset & $\mathcal{M}$ & Medium-confidence set\\
$\mathcal{T}$ & Training set & $\mathcal{L}$ & Low-confidence set\\
$\mathcal{V}$ & Validation set & $J_{\text{LSCE}}$ & Label Smoothing Cross-Entropy loss\\
$\hat{Y}$ & Predicted labels & $J_{\text{KLD}}$ & Kullback-Leibler Divergence loss\\
$Y$ & Real labels & $J_{\text{CE}}$ &  Cross-Entropy loss\\
\bottomrule
\end{tabular}
\label{symbols}
\end{table}

\subsection{Proposed model}
\label{learning_model}
\begin{figure}[htbp]
\centering  
\includegraphics[width=0.48\textwidth]{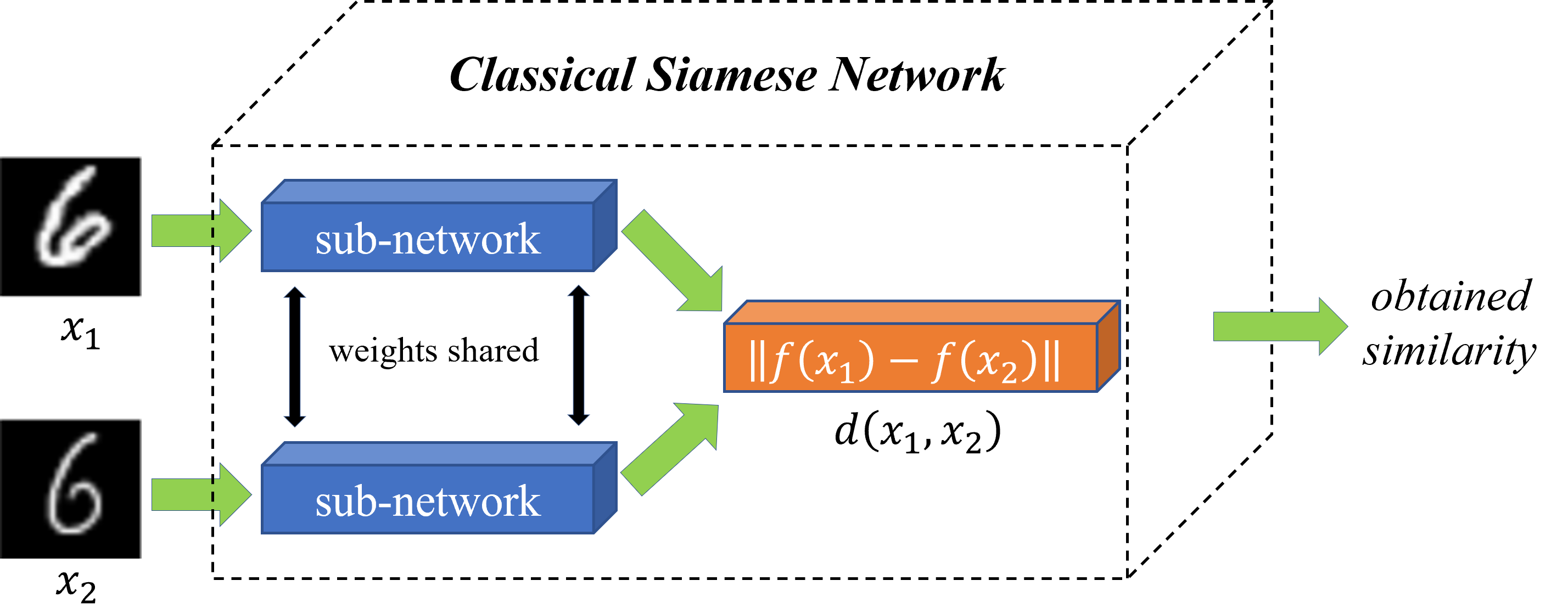}
\caption{The structure of the classical Siamese network.}
\label{siamese}
\end{figure}

Before introducing our proposed architecture, we provide a brief overview of the conventional Siamese network. Originally proposed in \cite{siamese_original} for comparing and verifying the similarity between two handwritten signatures, the classical Siamese network consists of a paired structure with two identical artificial neural networks, as illustrated in Fig. \ref{siamese}. This architecture takes two input samples and evaluates their similarity by computing the distance between their feature vectors, typically using Euclidean or Cosine distance. A key advantage of the Siamese network lies in its shared parameter framework, where the weights of the two sub-networks are identical, which not only reduces model complexity but also ensures that different inputs are mapped to the same feature space, maintaining consistent data distributions.

Though initially designed to measure the similarity between two inputs, the Siamese network also serves as an efficient feature extractor. Building upon this foundation, our proposed model, Siamese-GAP, integrates GAP modules at multiple levels to identify relevant features throughout the network. The Siamese-GAP model comprises a pair of parameter-sharing CNNs, with each sub-network partitioned into four blocks (comprising 2, 3, 4, and 4 layers, respectively) across 13 total layers. Each block includes a combination of convolutional layers with varying kernel sizes (32, 64, 128, and 256), Batch Normalization (BN), and Rectified Linear Unit (ReLU) activation layers. Given the importance of texture in medical imaging, our design prioritizes retaining spatial information from the input data. To achieve this, we replace pooling layers with strides for down-sampling, setting the stride of the first convolutional layer in each block to 2 from the second block onward while keeping it at 1 for the other blocks. After each block, a GAP layer is employed to derive a feature vector. These four feature vectors are concatenated to form a single GAP feature vector, capturing information across low to high levels within each sub-network. Notably, GAP was chosen over the traditional Global Max Pooling (GMP) for feature extraction due to its superior performance (refer to Section \ref{results} for detailed comparisons). GAP is calculated as follows: 

\begin{equation}
y_{k}=\frac{1}{\left | \mathbb{M}_k \right |}\sum_{(p,q)\in \mathbb{M}_k}^{}x_{kpq}
\end{equation}
where the single output value of the $k$-th feature map $\mathbb{M}$ produced using GAP is represented by $y_{k}$. $\left | \mathbb{M}_k \right |$ is the number of elements in the $k$-th feature map, and $x_{kpq}$ is the element at location $(p, q)$ of the $k$-th feature map.

The second sub-network in our Siamese-GAP model is structurally identical and similarly configured. The GAP-feature vectors derived from the two sub-networks are combined using an element-wise addition operation to produce the final feature vector, which is subsequently fed into a fully connected layer of size 480. Finally, a Softmax layer generates probability distributions for the KL-0 and KL-2 grades, predicting the final KOA grade. In addition to the element-wise addition, we evaluated the direct concatenation of the two GAP-feature vectors. However, this approach increased the number of parameters without yielding any performance improvement.

\subsection{Partition mechanism of the training set}
\label{Partition}
The objective of our proposed hybrid loss strategy is to assign a confidence level to each sample relative to its label, enabling more targeted and effective training. While a straightforward approach might involve directly dividing the training batch into subsets based on predefined confidence thresholds, the lack of prior knowledge about sample confidence levels in the original dataset necessitates an alternative approach. In our method, we assume that the high-confidence set ($\mathcal{H}$) comprises the majority of the dataset, reflecting samples that are likely to be classified correctly with greater certainty. Conversely, the low-confidence set ($\mathcal{L}$) represents the smallest portion of the dataset, encompassing samples with higher uncertainty or ambiguity. Such assumption aligns with the expectation for a meaningful dataset. On the other hand, samples that do not fall into either $\mathcal{H}$ or $\mathcal{L}$ are assigned to the medium-confidence set ($\mathcal{M}$), representing an intermediate level of classification confidence. The confidence level of each sample is determined based on the conditional probability of it being correctly classified, which is computed as:

\begin{equation}
c_s = P_{\hat{Y}|T} (\hat{Y} = k|Y = k), \quad \forall k \in \mathcal{K}, \forall s \in \mathcal{D}
\end{equation}
where $\hat{Y}$ and $Y$ represent the predicted and actual labels, respectively, $k$ denotes the KL grade of the sample $s$, $P_{\hat{Y}|T}$ represents the conditional probability distribution, $\mathcal{K}$ is the set of KL grades, and $\mathcal{D}$ refers to the overall dataset.

To determine the confidence subsets within each training batch, we first grouped the samples by their KL grade and ranked them in descending order based on their calculated confidence levels. Then, we introduced a hyper-parameter ratio, $\lambda$, to partition the batch into three subsets based on their confidence scores in a defined proportion: high-confidence ($\mathcal{H}$), medium-confidence ($\mathcal{M}$), and low-confidence ($\mathcal{L}$) sets. The same ratio $\lambda$ was applied uniformly to both KL-0 and KL-2 classes to ensure consistency across the dataset. Thus, each training batch was divided into the following subsets for KL-0 and KL-2:
\begin{itemize}
\item High-confidence: $\mathcal{H}_0$ and $\mathcal{H}_2$
\item Medium-confidence: $\mathcal{M}_0$ and $\mathcal{M}_2$
\item Low-confidence: $\mathcal{L}_0$ and $\mathcal{L}_2$
\end{itemize}

Such partitioning allows the model to address samples with varying levels of difficulty systematically. Specifically, high-confidence samples ($\mathcal{H}$) provide stable gradients and reinforce the model’s ability to generalize, while low-confidence samples ($\mathcal{L}$) help the model focus on challenging or ambiguous cases. The medium-confidence set ($\mathcal{M}$) acts as a transitional group, balancing the contributions of the other two sets. To fully leverage the characteristics of each confidence subset, a tailored hybrid loss strategy is employed.

\subsection{Hybrid loss strategy}
\label{hyperparaters}
Firstly, the high-confidence set, $\mathcal{H}$, is typically characterized by instances where the model’s predicted probabilities are close to 0 or 1. In this context, label smoothing \cite{LS}, a regularization technique, proves beneficial as it helps mitigate the risk of the model becoming overly confident in its predictions. By softening the predicted probabilities, label smoothing reduces the likelihood of the model making incorrect predictions due to noise or other factors, even when it exhibits high confidence in its outputs. To implement label smoothing, the traditional one-hot encoded label vector $y_s^{hot}$ is replaced with a smoothed version, referred to as $y_s^{LS}$, which represents a weighted mixture of $y_s^{hot}$:

\begin{equation}
y_{s}^{LS} = y_{s}^{hot}(1-\varepsilon)+ \frac{\varepsilon}{2}, \quad y_{s}^{hot} \in \left\{0,1\right\}, \varepsilon \in (0,1)
\end{equation}

\begin{equation}
y_{s}^{hot}=\left\{
\begin{array}{lcr}
1       &      & {\hat Y      =      Y}\\
0    &      & {\hat Y  \neq  Y}\\
\end{array} \right.
\end{equation}
where $y_{s}^{hot}$ denotes the one-hot encoded ground-truth label of the sample $s$, and $\varepsilon$ is a hyper-parameter that governs the degree of smoothing.

Following this, the LSCE loss, $J_{\text{LSCE}}$ is computed as follows:

\begin{equation}
\begin{aligned}
J_{\text{LSCE}} &= \sum_{s\in \mathcal{H}_k}^{}-y_{s}^{LS}log(c_s)\\&=\sum_{s\in \mathcal{H}_k}^{}-(y_s^{hot}(1-\varepsilon) + \frac{\varepsilon}{2})log(c_s),{\forall}k \in \mathcal{K}, {\forall}s \in \mathcal{H}
\end{aligned}
\label{LSCE}
\end{equation}

Secondly, for samples in the medium-confidence set $\mathcal{M}$, where the model’s predicted probabilities are neither close to 0 nor 1, indicating a degree of uncertainty, the traditional Cross-Entropy (CE) loss may fail to effectively address this ambiguity, leading to suboptimal training. To address this, the Kullback-Leibler Divergence (KLD) loss is employed to encourage the model to align its predictions with a target probability distribution that reflects the inherent uncertainty. Since the dataset was randomly divided during preprocessing, the confidence distribution of the validation set is assumed to be representative of the average confidence distribution of overall data. In this study, the target probability distribution for each KL grade is derived from the corresponding average confidence values obtained from the validation set (refer to Eq. \ref{p_target} and Eq. \ref{p_s}). The average confidence $c_k$ of each KL grade is computed as follows:

\begin{equation}
c_k = \frac{1}{|\mathcal{V}_k|} \sum_{s \in \mathcal{V}_k} c_s, \quad \forall k \in \mathcal{K}
\end{equation}
where $|\mathcal{V}_k|$ denotes the cardinality of the validation set $\mathcal{V}_k$.\\

The KLD loss, $J_{\text{KLD}}$ is thus given by:

\begin{equation}
J_{\text{KLD}} = \sum_{s\in \mathcal{M}_k}^{}y_s^{hot}log\frac{p_{target}}{p_s}, \quad {\forall}k \in \mathcal{K}, {\forall}s \in \mathcal{M}
\end{equation}
where, 
\begin{equation}
\label{p_target}
p_{target}=\left\{
\begin{array}{lcr}
c_k       &      & {\hat Y      =      Y}\\
1 - c_k     &      & {\hat Y  \neq  Y}\\
\end{array} \right.
\end{equation}
and 
\begin{equation}
\label{p_s}
p_s=\left\{
\begin{array}{lcr}
c_s       &      & {\hat Y      =      Y}\\
1 - c_s     &      & {\hat Y  \neq  Y}\\
\end{array} \right.
\end{equation}

Thirdly, for the low-confidence set $\mathcal{L}$, the model’s predicted probabilities are often distributed across multiple classes or resemble a uniform distribution, indicating significant uncertainty and a lack of clear indication of the correct class. In such cases, assigning a specific target label is inherently challenging. The traditional CE loss is particularly well-suited for these scenarios, as it evaluates the dissimilarity between the predicted probabilities and the target labels. By using the true labels as targets, even when they are uncertain or ambiguous, the CE loss directs the model to adjust its parameters to minimize the discrepancy between its predictions and the provided targets. Despite the uncertainty in the predicted probabilities, the true labels serve as a reliable reference, allowing the model to focus on learning the most discriminative features and improving its classification performance. The CE loss $J_{\text{CE}}$ is computed as follows:

\begin{equation}
J_{\text{CE}} = \sum_{s\in \mathcal{L}_k}^{}-y_{s}^{hot}log(c_s), \quad {\forall}k \in \mathcal{K}, {\forall}s \in \mathcal{L}
\end{equation}

Finally, the proposed hybrid loss function consists of combining three function losses: LSCE, KLD, and CE for the sets $\mathcal{H}$, $\mathcal{M}$ and $\mathcal{L}$, respectively, and is defined as:

\begin{equation}
J_{\text{hybrid}} = \alpha J_{\text{LSCE}} + \beta J_{\text{KLD}} + \gamma J_{\text{CE}}
\end{equation}
where the hyper-parameters $\alpha$, $\beta$, and $\gamma$ were employed to achieve a more balanced and weighted combination of the loss functions under consideration, which will be discussed in Section \ref{Selection}.

\section{Experiment settings}

\subsection{Employed knee database}
The Osteoarthritis Initiative (OAI) \cite{OAI} is a publicly available, large-scale, multi-centre longitudinal study designed to improve understanding of KOA. Established in 2002 with funding from the National Institutes of Health (NIH) and private industry partners, the OAI serves as a comprehensive resource for researchers seeking to advance the diagnosis, prevention, and treatment of KOA. The OAI includes a diverse cohort of 4,796 participants aged 45–79, with varying levels of risk for KOA. Over an extended follow-up period, participants undergo regular clinical assessments, imaging studies (e.g., X-rays and MRIs), and collection of biological samples, along with detailed surveys capturing lifestyle and health-related information.

\subsection{Data preprocessing}
\begin{figure}[htbp]
\centering
\subfloat[]{
\begin{minipage}[t]{0.195\textwidth}
\centering
\includegraphics[width=0.95\textwidth]{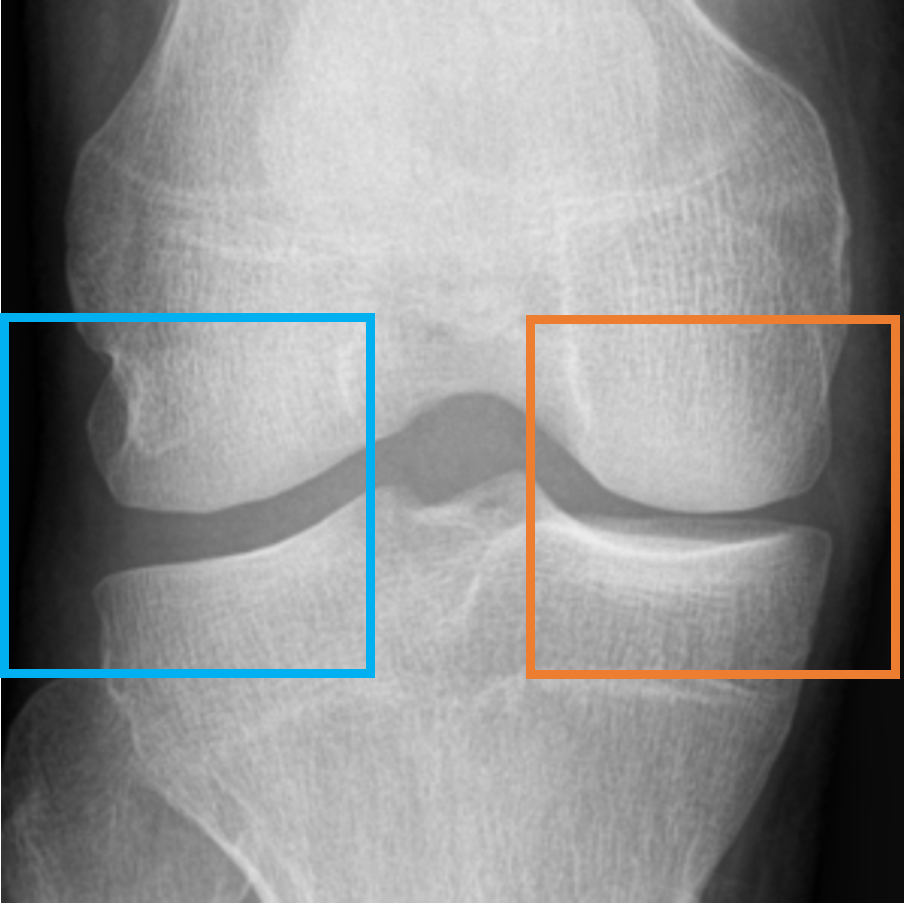}
\end{minipage}
}
\subfloat[]{
\begin{minipage}[t]{0.122\textwidth}
\centering
\includegraphics[width=0.75\textwidth]{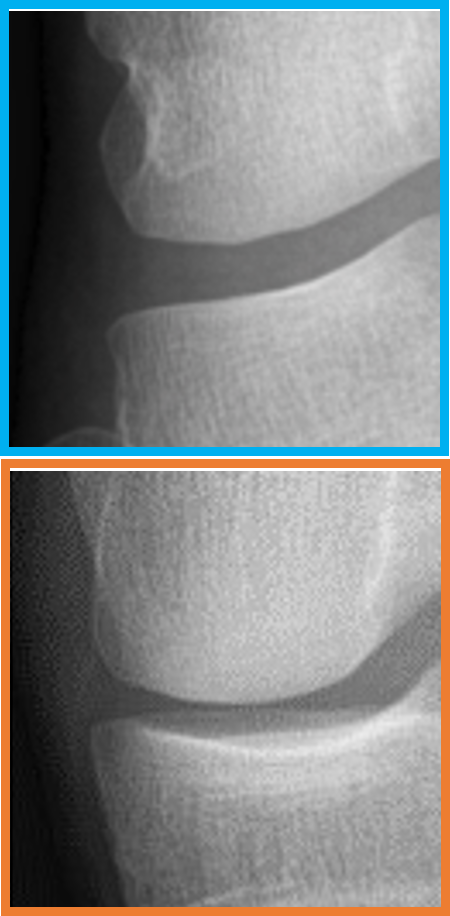}
\end{minipage}
}
\caption{Lateral and medial patches are in blue and orange boxes (a). Obtained patches serve as input of the model (b).}
\label{Fig-Patches}
\end{figure}

As shown in Fig. \ref{Fig-Patches}, two patches were cropped through a fully automated segmentation approach to encompass the entire KOA-related region, including osteophytes and JSN. Specifically, two 128 $\times$ 128 pixel patches were extracted from the lateral and medial one-third regions of the knee joint. To maintain consistency, the medial patch was flipped horizontally. These two patches were then used as inputs for our proposed model.

After preprocessing, a total of 3,185 KL-0 and 2,126 KL-2 images were obtained. The dataset was subsequently split into training, validation, and test sets in a 7:1:2 ratio, stratified by KL grade.

\subsection{Experimental details}
The weights of the network were initialized by the Kaiming initialization procedure \cite{kaiming}. With a learning rate of 1e-03 and a mini-batch size of 32, the Adam optimizer \cite{adam} was used to train 500 epochs. Random brightness, rotation, and gamma correction were executed randomly in the data augmentation process. To mitigate over-fitting, weight decay was implemented with a coefficient of 3e-04, while dropout was applied at a rate of 0.2. Moreover, to deal with the imbalanced dataset, the set of KL-2 was oversampled based on bootstrapping \cite{oversample}. The implementation of our work was performed using Nvidia TESLA A100 GPUs, each equipped with 80GB of RAM, and PyTorch v1.8.1 \cite{pytorch}.

\section{Results and discussion}
In this study, KL-2 was designated as the positive class, reflecting our classification goal of accurately identifying instances of KL-2. Additionally, to provide an independent evaluation of our proposed model, the accuracy values presented in Table \ref{ablation} and Table \ref{comparision} were calculated using the traditional CE loss function, ensuring a consistent baseline for performance comparison.

\subsection{Ablation study for GAP/GMP modules}
\label{results}
In our proposed learning model, in addition to the final regular GAP layer, three additional GAP layers were integrated into each sub-network at different positions (P1, P2, P3, and P4) to capture multi-level features effectively. The placement of these GAP layers determines the granularity and significance of the features extracted from different depths of the network. To evaluate the contribution of each GAP layer, we conducted an ablation study comparing the performance of GAP and GMP units at various combinations of positions, as summarized in Table \ref{ablation}. As can be seen, the results consistently show that configurations leveraging GAP layers outperform those using GMP layers across all combinations. For instance, replacing GMP with GAP at P2 improves accuracy from 87.86$\%$ to 88.24$\%$, and incorporating GAP at all positions (P1, P2, P3, and P4) achieves the highest accuracy of 88.38$\%$, compared to 86.95$\%$ with GMP. This superior performance of GAP can be attributed to its ability to preserve the average value of feature maps, which effectively retains global texture and contextual information. In contrast, GMP focuses solely on the maximum value of each feature map, discarding substantial texture-related and fine-grained details, thereby limiting its representational power. Such observation aligns with findings reported in \cite{gapvsgmp}. The incremental improvements observed as GAP units are added at different positions highlight the importance of multi-level feature aggregation in our proposed model. For example, adding GAP at P2 (in addition to P4) increases accuracy from 87.78$\%$ to 88.24$\%$, underscoring the utility of incorporating intermediate-level features into the final representation. Furthermore, the combination of GAP layers at P1, P2, P3, and P4 achieves the best performance, emphasizing the complementary nature of features extracted at varying depths of the network. By leveraging GAP layers effectively, our model maximizes the utility of texture and contextual information, resulting in improved performance and robustness.

\begin{table}[htbp]
\caption{Contribution of GAP and GMP units}
\setlength{\tabcolsep}{4mm}
\label{ablation}
\centering
\begin{threeparttable}
\begin{tabular}{ccccccc} 
\toprule
\multicolumn{4}{c}{\bf Position$^1$}  & & \multicolumn{2}{c}{\bf Accuracy ($\%$)} \\
\cmidrule(lr){1-4}  
\cmidrule(lr){6-7}  
P1 & P2 & P3 & P4 && GAP & GMP\\
\midrule
 &   &  & $\times$  && 87.51 & 87.12\\ 
$\times$ &   &  & $\times$  && 88.01 & 87.94\\
     &   $\times$&  & $\times$  && 88.24 & 87.86\\ 
  &   & $\times$ & $\times$  &&  87.78 & 87.90\\ 
$\times$ & $\times$  & & $\times$ && 88.14 & 86.89\\
$\times$ & $\times$  & $\times$ & $\times$  && \bf 88.38 & 86.95\\
\bottomrule
\end{tabular}
\begin{tablenotes}
\footnotesize
\item[$1$] P1, P2, P3 and P4 refer to positions of the GAP/GMP unit in our proposed network.
\end{tablenotes}
\end{threeparttable}
\end{table}

\subsection{Comparison with other learning models}
We evaluated the performance of our proposed Siamese-GAP model against several widely used CNN architectures and State-Of-The-Art (SOTA) methods, as summarized in Table \ref{comparision}. To ensure a fair comparison, all models were trained and tested using the same knee joint input data. The metrics used for evaluation included accuracy (Acc), F1-score (F1), and the number of model parameters (Params). Our model achieved the highest accuracy (88.38$\%$) and F1-score (85.93$\%$) among all compared models, demonstrating its superior predictive performance in assessing KOA. Specifically, among the traditional CNNs, DenseNet-201 (84.43$\%$ accuracy, 81.29$\%$ F1-score) and ResNet-18 (83.59$\%$ accuracy, 80.30$\%$ F1-score) performed relatively well. However, these models required significantly more parameters, with DenseNet-201 using 18.09 million and ResNet-18 using 11.17 million parameters. Compared to SOTA models, our model still demonstrated substantial improvements in both accuracy and F1-score while maintaining a lightweight design. Compared to Tiulpin et al. \cite{tiuplin}, though our model incurs a modest increase in parameter count by 0.35 million, this slight increase is accompanied by significant performance improvements, with accuracy rising by 1.05$\%$ and F1-score by 1.11$\%$. It is noteworthy that these advancements are particularly impactful in the context of early KOA assessment, where even small improvements in predictive performance can translate into meaningful clinical benefits, enhancing diagnostic accuracy and decision-making processes.

\begin{table}[htbp]
\centering
\caption{Comparison of common and SOTA models}
\setlength{\tabcolsep}{4mm}
\begin{threeparttable}
\begin{tabular}{lccc} 
\toprule
Model & Acc ($\%$)  & F1 ($\%$) & Params ($M$)$^{1}$\\
\midrule
Densenet-121 & 80.84 & 77.03 & 6.95\\
Densenet-161 & 79.06 & 75.09 & 26.47\\
Densenet-169 & 81.81 & 78.23 & 12.48\\
Densenet-201 & 84.43 & 81.29 & 18.09\\
Resnet-18 & 83.59 & 80.30 & 11.17\\
Resnet-34 & 80.14 & 76.35 & 21.28\\
Resnet-50 & 81.78 & 78.29 & 23.51\\
Resnet-101 & 77.60 & 72.98 & 42.50\\
Resnet-152 & 76.47 & 72.13 & 58.14\\
VGG-11 & 80.97 & 77.30 & 195.89\\
VGG-13 & 80.71 & 77.88 & 196.07\\
VGG-16 & 73.17 & 68.50 & 201.38\\
VGG-19 & 71.98 & 68.84 & 206.70\\
Nasser et al. \cite{yassine} & 83.40 & 84.34 & 11.37\\
Antony et al. \cite{antony1} & 84.50 & 81.31 & 5.40\\
Tiulpin et al. \cite{tiuplin} & 87.33 & 84.82 & \bf 2.36\\
Ours & \bf 88.38 & \bf 85.93 & 2.71\\
\bottomrule
\end{tabular}
\begin{tablenotes}
\footnotesize
\item[$1$] Params is the number of parameters employed.
\end{tablenotes}
\end{threeparttable}
\label{comparision}
\end{table}

\begin{figure}[htbp]
\centering
\subfloat[KL-2 knee joint]{
\begin{minipage}[t]{0.145\textwidth}
\centering
\includegraphics[width=1\textwidth]{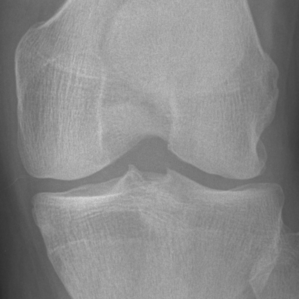}
\end{minipage}
}
\subfloat[DenseNet-201]{
\begin{minipage}[t]{0.145\textwidth}
\centering
\includegraphics[width=1\textwidth]{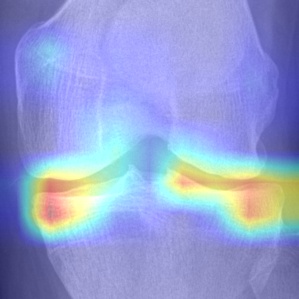}
\end{minipage}
}
\subfloat[ResNet-18]{
\begin{minipage}[t]{0.145\textwidth}
\centering
\includegraphics[width=1\textwidth]{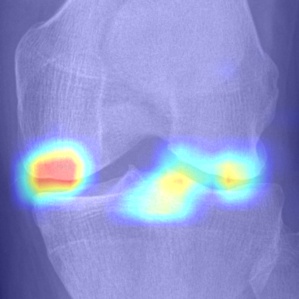}
\end{minipage}
}
\quad
\subfloat[VGG-11]{
\begin{minipage}[t]{0.145\textwidth}
\centering
\includegraphics[width=1\textwidth]{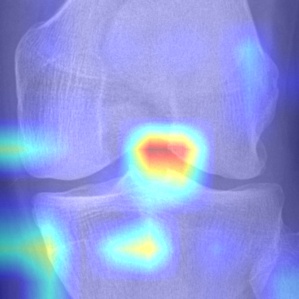}
\end{minipage}
}
\subfloat[Tiulpin et al. \cite{tiuplin}]{
\begin{minipage}[t]{0.145\textwidth}
\centering
\includegraphics[width=1\textwidth]{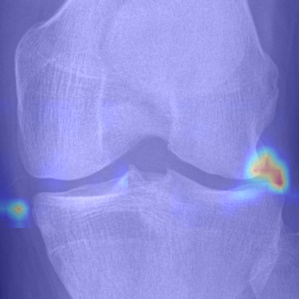}
\end{minipage}
}
\subfloat[Ours]{
\begin{minipage}[t]{0.145\textwidth}
\centering
\includegraphics[width=1\textwidth]{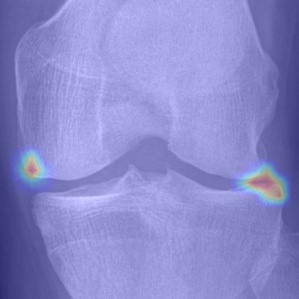}
\end{minipage}
}
\caption{Comparison of the attention maps of the evaluated models.}
\label{Attention-Map}
\end{figure}

To gain a deeper understanding of the model's decision-making process, the Grad-CAM technique \cite{gradcam} was employed to generate heatmaps, highlighting the regions of interest that the models attend to during classification. For clarity and conciseness, we present heatmaps for only the best-performing models within each architectural series as well as the SOTA methods. As shown in Fig. \ref{Attention-Map}, all evaluated models demonstrated some degree of responsiveness to regions indicative of early KOA features. However, many models also exhibited significant attention to irrelevant background regions. Such misdirected attention not only dilutes the feature extraction process but also increases the likelihood of misclassification, undermining the overall performance of the model. In contrast, our proposed model stands out by focusing more precisely and consistently on regions directly associated with KOA, such as joint spaces and subchondral bone areas. This targeted attention improves the extraction of discriminative features, which directly enhances classification performance. Furthermore, our approach demonstrates a marked reduction in attention to irrelevant regions, reflecting a more refined and robust understanding of the radiographic features specific to early KOA. Notably, when compared to Tiulpin et al. \cite{tiuplin}, our method exhibits superior localization of KOA-affected areas, with heatmaps showing concentrated and precise attention on pathological regions, which underscores the effectiveness of our model in capturing the subtle features indicative of early KOA. By leveraging this enhanced capability, our proposed framework not only achieves higher classification accuracy but also demonstrates increased interpretability and reliability.

\subsection{Settings of the hyper-parameters}
\label{Selection}
While the previous section highlights the superior predictive capabilities of the our model, such performance can be further enhanced by combining the hybrid loss strategy. To evaluate the impact of hyper-parameters in the proposed hybrid loss strategy, we tested various configurations. First, we explored different smoothing levels $\varepsilon \in \{0.05, 0.1, 0.15, 0.2\}$. Next, based on the assumption outlined in Section \ref{Partition}, we explored various partition ratios $\lambda$, specifically 6:3:1, 7:2:1, and 8:1:1, opting for whole integer values to simplify implementation. Subsequently, we conducted a grid search to determine the optimal weight hyper-parameters ($\alpha$, $\beta$, $\gamma$) within the range $[0.1, 1]$ while ensuring that $\alpha + \beta + \gamma = 1$. For simplicity, Table \ref{abc} presents only the performance results for each $\lambda$ and $\varepsilon$ configuration, using the optimized weight parameters $\alpha$, $\beta$, and $\gamma$. As can be seen, the best performance was achieved with a partition ratio of $\lambda = 7:2:1$, $\varepsilon = 0.15$, $\alpha = 0.4$, $\beta = 0.5$, and $\gamma = 0.1$. Such configuration is employed in all subsequent experiments throughout the following sections.

Notably, the results reveal several key trends. First, among the tested partition ratios, $\lambda = 7:2:1$ consistently delivered the best performance, suggesting that this ratio may closely reflect the natural distribution of confidence levels across subsets within the original dataset. Secondly, a moderate smoothing level ($\varepsilon = 0.15$) produced the best results, which may reflect an average confidence level of approximately 85$\%$ in the medium-confidence set. Finally, the optimized weights ($\alpha = 0.4, \beta = 0.5$) underscore the balanced importance of high- and medium-confidence samples in guiding the learning process. Meanwhile, the significantly lower weight assigned to low-confidence samples ($\gamma = 0.1$) emphasizes the critical necessity of suppressing their influence, thereby preventing the model from being adversely affected by potentially noisy or unreliable data.

\begin{table}[htbp]
\centering
\caption{Performance evaluation under various hyper-parameter configurations}
\setlength{\tabcolsep}{3.4mm}
\begin{tabular}{ccccccc} 
\toprule
$\lambda$ & $\varepsilon$ & $\alpha$  & $\beta$  & $\gamma$ & Acc ($\%$) & F1 ($\%$)\\
\midrule
\multirow{4}{*}{6:3:1} & 0.05 & 0.4 & 0.5 & 0.1 & 88.64 & 86.15\\
& 0.1 & 0.4 & 0.4 & 0.2 &  88.59 & 86.08\\
& 0.15 & 0.5 & 0.4 & 0.1 &  88.75 & 86.28\\
& 0.2 & 0.3 & 0.5 & 0.2 & 88.82 & 86.39\\
\midrule
\multirow{4}{*}{\underline{7:2:1}} & 0.05 & 0.3 & 0.5 & 0.2 & 88.96 & 86.59\\
& 0.1 & 0.4 & 0.3 & 0.3 &  88.51 & 86.02\\
& \underline{0.15} & \underline{0.4} & \underline{0.5} & \underline{0.1} & \bf 89.14 & \bf 86.78\\
& 0.2 & 0.3 & 0.5 & 0.2 & 88.82 & 86.34\\
\midrule
\multirow{4}{*}{8:1:1} & 0.05 & 0.3 & 0.5 & 0.2 & 89.09 & 86.67\\
& 0.1 & 0.4 & 0.4 & 0.2 &  88.56 & 86.02\\
& 0.15 & 0.3 & 0.4 & 0.3 &  88.67 & 86.10\\
& 0.2 & 0.3 & 0.4 & 0.3 & 89.03 & 86.50\\
\bottomrule
\end{tabular}
\label{abc}
\end{table}

\subsection{Contribution of the hybrid loss strategy}
\begin{figure}[htbp]
\centering
\subfloat[with the single CE loss]{
\includegraphics[width=0.235\textwidth]{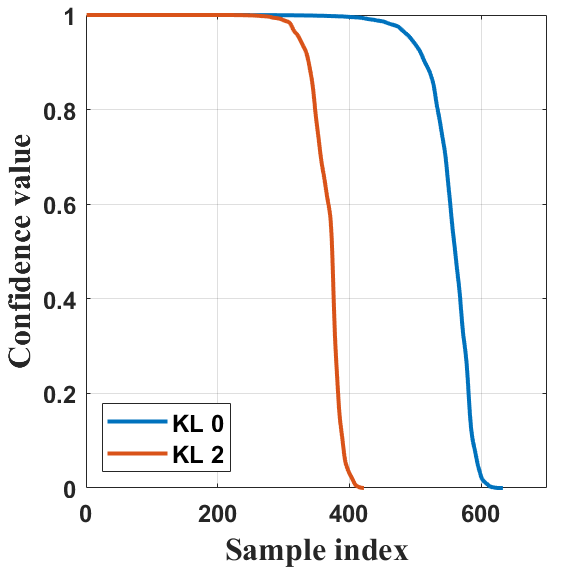}
\label{confidence_ce5}}
\subfloat[with the hybrid loss]{
\includegraphics[width=0.235\textwidth]{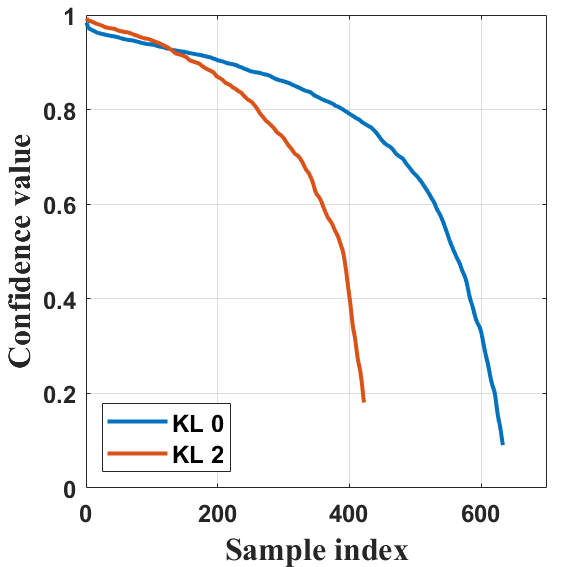}
\label{confidence_hybrid}}
\caption{Comparison of confidence distributions. Subfigure (a) shows the distribution under the single CE loss function, while subfigure (b) illustrates the distribution using the hybrid loss strategy. The confidence scores are sorted in descending order.}
\label{confidence_disctibution}
\end{figure}

As illustrated in Fig. \ref{confidence_disctibution}, the hybrid loss strategy demonstrates its effectiveness by significantly mitigating overconfidence in the model compared to the traditional CE loss function, as observed on the same test set. As shown in Fig. \ref{confidence_ce5}, the confidence distribution obtained using the traditional single CE loss exhibits sharp transitions and overly concentrated values near 1 for both classes. This overconfidence not only reduces inter-class continuity but also risks impairing the model's generalization ability. In contrast, Fig. \ref{confidence_hybrid} show that the hybrid loss strategy produces a smoother and more evenly distributed confidence curve across classes, which ensures better inter-class continuity and adjusts the confidence levels for each sample more appropriately. Furthermore, the hybrid loss strategy minimizes the likelihood of excessive confidence increases. The smoother transitions and balanced confidence levels foster better alignment between the model's predictions and the inherent variability in the data, which ultimately enhances robustness and stability during training.

\begin{figure}[htbp]
\centering
\subfloat[with the single CE loss]{
\begin{minipage}[t]{0.23\textwidth}
\centering
\includegraphics[width=1\textwidth]{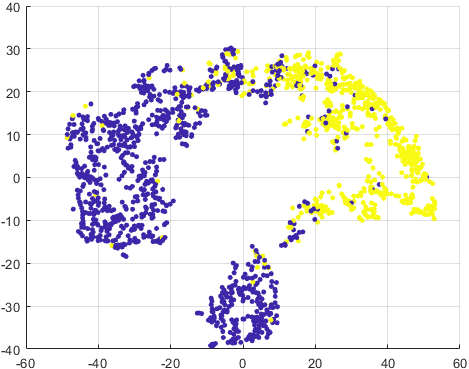}
\label{tnse_ce}
\end{minipage}
}
\subfloat[with the hybrid loss]{
\begin{minipage}[t]{0.23\textwidth}
\centering
\includegraphics[width=1\textwidth]{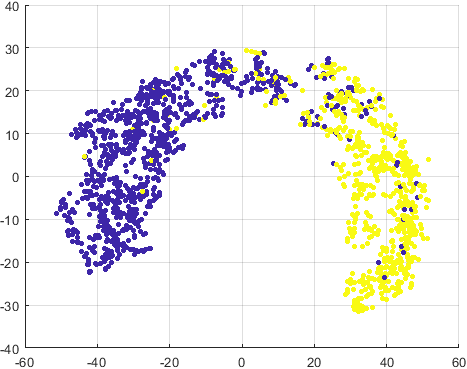}
\label{tnse_hybrid}
\end{minipage}
}
\caption{t-SNE scatter plots visualizing the feature representations. (a) depicts the results using the single CE loss, while (b) illustrates the results achieved with the hybrid loss strategy.}
\label{tnse}
\end{figure}

On the other hand, the t-SNE scatter plots presented in Fig. \ref{tnse} illustrate the advantages of the hybrid loss strategy, particularly for confidence-sensitive tasks like early KOA classification. Specifically, under the CE loss (Fig. \ref{tnse_ce}), while some degree of intra-class cohesion is observed, the clusters are relatively scattered with noticeable overlap, indicating limited inter-class separability. In contrast, the hybrid loss (Fig. \ref{tnse_hybrid}) produces significantly more compact clusters, thereby reducing intra-class variance. At the same time, it increases inter-class separability, with distinctly defined boundaries between different classes. This marked improvement in feature distribution not only demonstrates the hybrid loss strategy's ability to foster a more discriminative feature space but also emphasizes its role in enhancing overall classification reliability. Moreover, the reduced overlap and tighter clusters directly translate to fewer classification errors, which is particularly crucial in confidence-sensitive early KOA detection.

\subsection{Evaluation of the applicability of the hybrid loss strategy}
To further evaluate the applicability and effectiveness of the proposed hybrid loss strategy, we conducted a performance analysis using a diverse set of high-performing models. These models were trained with both the traditional CE loss and the proposed hybrid loss strategy under identical hyper-parameter configurations outlined in Section \ref{Selection}. The comparative results are summarized in Table \ref{applicability}. As can be seen, all evaluated models demonstrated accuracy improvements when trained with the hybrid loss strategy, underscoring the strategy's capability to enhance performance across diverse architectures. The magnitude of improvement varied by architecture. Specifically, ResNet-18 and VGG-11 exhibited substantial gains of 1.37$\%$ and 2.09$\%$, respectively, while Densenet-201 showed a notable improvement of 1.48$\%$, suggesting that the hybrid loss effectively addresses potential limitations in feature representation learning. For SOTA models, Nasser et al. \cite{yassine}, Antony et al. \cite{antony1} and Tiulpin et al. \cite{tiuplin} demonstrated accuracy improvements of 1.25$\%$, 1.51$\%$, and 0.80$\%$, respectively, validating the generalizability of the hybrid loss across models designed for varied methodologies. Among all the models, our proposed model achieved the highest overall accuracy, improving from 88.38$\%$ to 89.14$\%$. While the absolute improvement of 0.77$\%$ appears moderate, it is particularly significant given the already high baseline accuracy. Such a level of improvement is especially impactful in early KOA detection, where even small gains in accuracy can lead to improved diagnostic precision and better clinical outcomes. Moreover, it is noteworthy that a critical strength of the hybrid loss strategy lies in its effectiveness even when the hyper-parameters are not specifically optimized for each individual model. These findings collectively demonstrate the utility of the proposed hybrid loss strategy in diverse learning contexts, offering robust and meaningful enhancements.

\begin{table}[htbp]
\centering
\caption{Performance comparison of models under different training strategies}
\label{applicability}
\setlength{\tabcolsep}{4.2mm}
\begin{threeparttable}
\begin{tabular}{lccc} 
\toprule
Models & Acc ($\%$)$^1$  & Acc ($\%$)$^2$  & Diff ($\%$)\\
\midrule
Densenet-201 & 84.43 & 85.91 & 1.48 $\uparrow$\\
Resnet-18 & 83.59 & 84.96 & 1.37 $\uparrow$\\
VGG-11 & 80.97 & 83.06 & 2.09 $\uparrow$\\
Nasser et al. \cite{yassine} & 83.40 & 84.65 & 1.25 $\uparrow$\\
Antony et al. \cite{antony1} & 84.50 & 86.01 & 1.51 $\uparrow$\\
Tiulpin et al. \cite{tiuplin} & 87.33 & 88.13 & 0.80 $\uparrow$\\
Our model & \bf 88.38 & \bf 89.15 & 0.77 $\uparrow$\\
\bottomrule
\end{tabular}
\begin{tablenotes}
\footnotesize
\item[$1$] Accuracy obtained using the single CE loss function
\item[$2$] Accuracy obtained using the proposed hybrid loss strategy
\end{tablenotes}
\end{threeparttable}
\end{table}

\subsection{Comparison between radiologists and our approach}
\subsubsection{Statistical metrics-based comparison}
To validate the effectiveness of our proposed global approach, we compared its performance against the decisions made by experienced radiologists. Specifically, 300 samples (150 KL-0 and 150 KL-2) in the testing set were randomly selected and independently labelled by three senior radiologists. To evaluate the consistency between the model and the radiologists, Cohen’s kappa coefficient ($\kappa$) was calculated to measure agreement, and McNemar's test was performed to assess whether the differences between their decisions were statistically significant. As shown in Table \ref{radiologist_vs_model}, the model achieved an accuracy of 89.2$\%$, while the radiologists' accuracy ranged from 92.0$\%$ to 93.3$\%$, reflecting their better overall performance. The model’s sensitivity (84.7$\%$) and specificity (93.3$\%$) were lower than those of the radiologists, whose sensitivities ranged from 91.0$\%$ to 92.0$\%$ and specificities ranged from 93.0$\%$ to 94.7$\%$, demonstrating the radiologists’ slightly stronger ability to distinguish KL-2. The agreement between the model and radiologists, measured by Cohen’s kappa, was substantial, with $\kappa$ values ranging from 0.85 to 0.87 for comparisons between the model and individual radiologists. Additionally, McNemar's test yielded p-values of 0.53, 0.37, and 0.51 for comparisons between the model and the three radiologists, respectively. All p-values exceeded the threshold for statistical significance ($p > 0.05$), indicating no statistically significant differences between the model’s predictions and those of the radiologists. These findings suggest that the proposed global approach achieves a diagnostic performance level comparable to that of experienced radiologists. By maintaining high agreement with radiologists and effectively balancing sensitivity and specificity, the model demonstrates significant potential as an auxiliary tool to assist in the early detection of KOA, potentially reducing the workload of clinical experts and enhancing diagnostic accuracy.

\begin{table}[htbp]
\centering
\caption{Statistical comparison between radiologists and model}
\label{radiologist_vs_model}
\setlength{\tabcolsep}{2.7mm}
\begin{tabular}{lccccc}
\toprule
& Acc ($\%$) & Sens ($\%$) & Spec ($\%$) & $\kappa$ & p-value\\ 
\midrule
Rad 1   & 92.0  & 91.0 & 93.0 & 0.85 & 0.53 \\
Rad 2   & 93.3  & 92.0 & 94.7 & 0.87 & 0.37 \\
Rad 3   & 93.0  & 91.7 & 94.3 & 0.86 & 0.51 \\
Model   & 89.2  & 84.7 & 93.3 & -    & -   \\
\bottomrule
\end{tabular}
\end{table}

\subsubsection{Confidence distribution-based comparison}
\begin{figure}[htbp]
\centering
\subfloat[from the model]{
\includegraphics[width=0.231\textwidth]{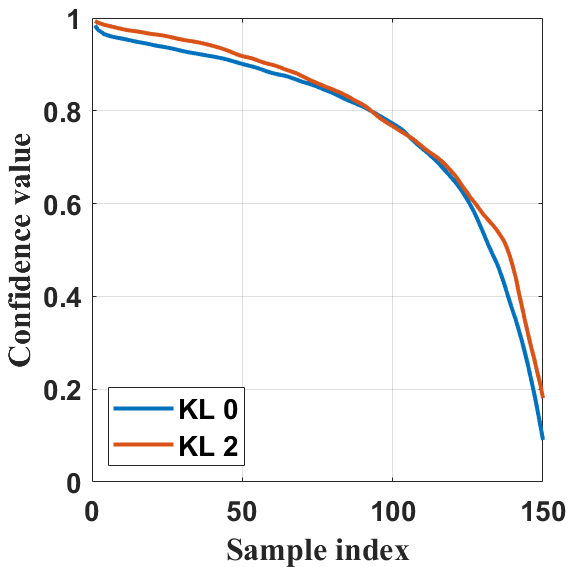}
\label{confidence_150_model}}
\subfloat[from the radiologists]{
\includegraphics[width=0.231\textwidth]{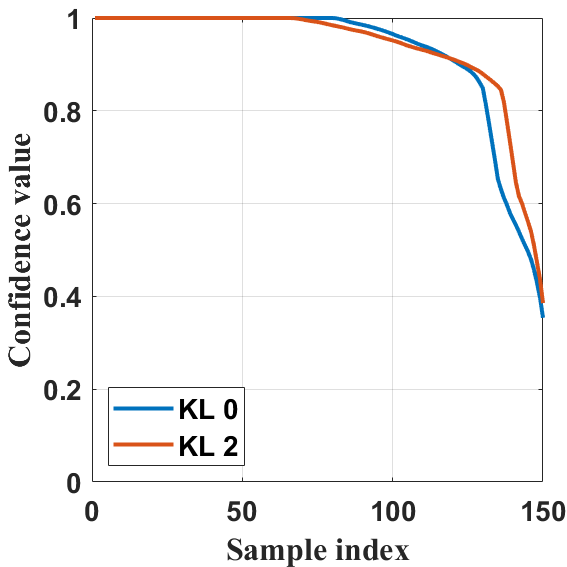}
\label{confidence_150_radiologist}}
\caption{Comparison of confidence distributions. Subfigure (a) represents the model's confidence distribution, while subfigure (b) illustrates the averaged confidence distribution of the radiologists. The confidence scores are sorted in descending order.}
\label{confidence_model_radiologist}
\end{figure}

Fig. \ref{confidence_150_model} and Fig. \ref{confidence_150_radiologist} illustrate the confidence curves for the model and the averaged radiologist predictions, respectively. As can be seen, both the model and radiologists exhibit similar confidence distributions across KL-0 and KL-2 samples, which suggests that both the model and radiologists rely on comparable decision-making processes when assigning confidence to classifications. Specifically, KL-2 samples consistently show higher confidence values compared to KL-0 samples, especially for correctly classified cases, which reflects the ability of both the model and radiologists to more confidently identify KL-2 cases. Furthermore, the overall trend of confidence decay from high certainty at the extremes to lower confidence near the boundary is preserved across both sets of predictions, reinforcing their alignment in judgment criteria. While the overall distributions are similar, some differences are notable. The model demonstrates a steeper decline in confidence near the boundary, which indicates the model’s sensitivity to borderline cases, where small variations in features significantly influence its decisions. In contrast, radiologists display a more gradual decline in confidence near the boundary, suggesting a cautious approach to borderline classifications, which reflects their expertise in dealing with uncertain cases.

Overall, these findings collectively demonstrate that the model not only achieves a high diagnostic accuracy close to that of radiologists but also aligns well with their confidence distributions, making it a reliable and interpretable auxiliary tool for early KOA detection.

\subsection{Rethinking KOA full/five-grade classification}
\begin{figure}[htbp]
\centering
\subfloat[Full/Five-grade classification]{
\begin{minipage}[t]{0.23\textwidth}
\centering
\includegraphics[width=1\textwidth]{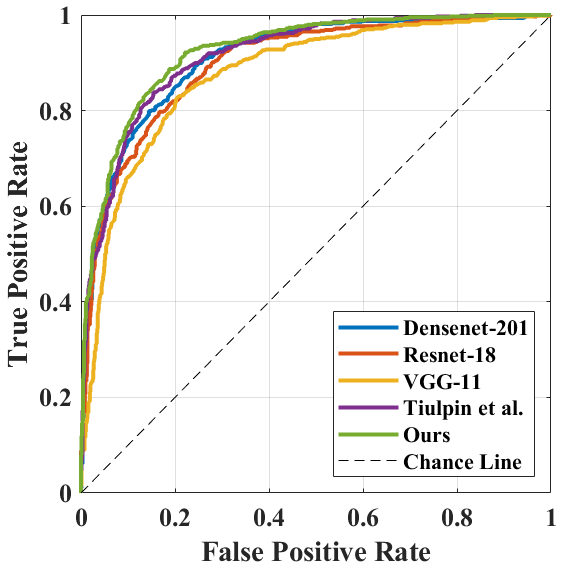}
\label{AOC1}
\end{minipage}
}
\subfloat[Binary classification]{
\begin{minipage}[t]{0.23\textwidth}
\centering
\includegraphics[width=1\textwidth]{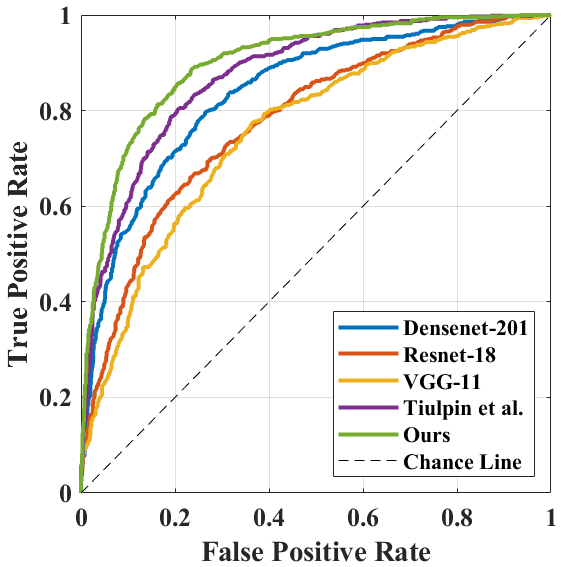}
\label{AOC2}
\end{minipage}
}
\caption{ROC curves for full/five-grade classification (KL$<$2 vs. KL$\geq$2) (a) and binary classification (KL-0 vs. KL-2) (b).}
\label{aoc}
\end{figure}

While our study primarily focuses on the clinically significant task of detecting early KOA stages KL-0 and KL-2, we also reassessed the model's performance across all KL grades to ensure a rigorous analysis. The evaluation was conducted by comparing our model with several SOTA models using the Area Under the Curve (AUC) as the primary metric. As shown in Fig. \ref{AOC1}, for the full/five-grade classification task, while our model demonstrated superior performance with the highest AUC of 0.92, the differences in AUC between our model and other SOTA models were relatively small, with DenseNet-201 and Tiulpin et al. achieving AUCs of 0.91, which can be attributed to the distinct and easily identifiable traits of advanced KOA stages (i.e., KL-3 and KL-4), such as pronounced joint space narrowing and large osteophyte formation. These prominent features result in detection accuracies that typically exceed 95$\%$ for most models, reducing the differentiation in AUC values among the evaluated models. On the other hand, for the binary classification task focused on the early detection of KOA (i.e., KL-0 vs. KL-2), as shown in Fig. \ref{AOC2}, our model achieved the highest AUC of 0.90, demonstrating a clear improvement over other evaluated models, including Tiulpin et al. (0.87). Such improvement is particularly noteworthy given the subtle visual distinctions in early KOA stages, where features such as minor JSN and incipient osteophyte formation are more nuanced and challenging to identify. Overall, these results demonstrate that while the performance of most models is comparable for advanced KOA stages, our model stands out for its robustness and accuracy in detecting early KOA, which is crucial in addressing a key clinical need.

\subsection{Discussion}
This study introduced a novel Siamese-GAP network combined with a hybrid loss strategy for early detection of KOA, specifically distinguishing KL-0 from KL-2 grades. By integrating multi-level feature extraction through GAP layers and a confidence-driven loss partitioning mechanism, the proposed framework addresses critical challenges in both feature representation and training sample variability. Specifically, the Siamese-GAP model demonstrated its superior feature extraction capabilities compared to traditional architectures and SOTA models. The integration of multi-level GAP layers enabled the retention of fine-grained and high-level features, enhancing the classification of subtle differences between KL-0 and KL-2 stages. The hybrid loss strategy, tailored to handle confidence variability in annotations, further improved the robustness and adaptability of the model. By partitioning training samples into high-, medium-, and low-confidence subsets and applying distinct loss functions, the model effectively learned from diverse data distributions while mitigating overconfidence issues observed in traditional CE loss. In comparison with radiologists, the model achieved diagnostic performance that closely matched expert-level accuracy, sensitivity, and specificity. Cohen’s kappa values indicated substantial agreement, and the results of McNemar's test confirmed no statistically significant differences between the model and radiologists’ predictions, which highlights the model's potential as an auxiliary tool, providing consistent and reliable diagnoses while reducing the workload of clinical experts. This is further validated by the confidence curves, which revealed that both the model and radiologists exhibited similar decision-making patterns.

Despite its strengths, the study has certain limitations. The hybrid loss strategy's reliance on manually tuned hyper-parameters presents computational challenges. Future work should explore automated hyper-parameter optimization and dynamic confidence partitioning mechanisms. Additionally, as this study is based on publicly available datasets, further validation on private hospital datasets is necessary to enhance its applicability and robustness for downstream deployment.

\section{Statement}
This manuscript was prepared using data from the OAI. The views expressed in it are those of the authors and do not necessarily reflect the opinions of the OAI investigators, the National Institutes of Health (NIH), or the private funding partners.

\section{Acknowledgments}
The authors would like to express their heartfelt appreciation to the French National Research Agency (ANR) for supporting this work through the ANR-20-CE45-0013-01 project.

\bibliographystyle{IEEEtran}
\bibliography{refs}

\end{document}